# An LSB Data Hiding Technique Using Natural Numbers

Sandipan Dey (1), Ajith Abraham (2), Sugata Sanyal (3)

[1] *Anshin Software Private Limited, Kolkata – 700091*

[2] *Centre for Quantifiable Quality of Service in Communication Systems*
*Norwegian University of Science and Technology, Norway*

[3] *School of Technology and Computer Science, Tata Institute of Fundamental Research, India*

*sandipan.dey@gmail.com, ajith.abraham@ieee.org, sanyal@tifr.res.in*

## Abstract

*In this paper, a novel data hiding technique is proposed, as an improvement over the Fibonacci LSB data-hiding technique proposed by Battisti et al. [1] based on decomposition of a number (pixel-value) in sum of natural numbers. This particular representation again generates a different set of (virtual) bit-planes altogether, suitable for embedding purposes. We get more bitplanes than that we get using Prime technique [2]. These bitplanes not only allow one to embed secret message in higher bit-planes but also do it without much distortion, with a much better stego-image quality, and in a reliable and secured manner, guaranteeing efficient retrieval of secret message. A comparative performance study between the classical Least Significant Bit (LSB) method, the Fibonacci LSB data-hiding technique and the proposed schemes indicate that image quality of the stego-image hidden by the technique using the natural decomposition method improves drastically against that using Prime and Fibonacci decomposition technique. Experimental results also illustrate that, the stego-image is visually indistinguishable from the original cover-image. Also we show the optimality of our technique.*

## 1. Introduction

Data hiding technique is a new kind of secret communication technology. While cryptography scrambles the message so that it can't be understood, steganography hides the data so that it can't be observed. In this paper, we discuss about a new decomposition method for classical LSB data-hiding technique, in order to make the technique more secure and hence less predictable. We generate a new set of (virtual) bit planes using our decomposition technique and embed data bit in these bit planes.

The Fibonacci LSB Data Hiding Technique proposed by Battisti et al. [1] investigates decomposition into a different set of bit-planes, based on the Fibonacci–*p*-sequences, given by,

$$F_p(0) = F_p(1) = 1$$

$$F_p(n) = F_p(n-1) + F_p(n-p-1), \forall n \geq 2, n \in \mathrm{N}$$

and embed a secret message-bit into a pixel if it passes the Zeckendorf condition, then during extraction, follow the reverse procedure.

We proposed the data hiding technique using prime decomposition [2] as an improvement over Fibonacci. Virtual bit-planes are generated using Prime Decomposition. The weight function of the Prime Number System is defined as:

$$P(0) = 1, P(i) = p_i, \forall i \in Z^+, p_i = i^{th} \operatorname{Prime},$$

$$p_0 = 1, p_1 = 2, p_2 = 3, p_3 = 5, ..$$

and embed a secret message-bit into a pixel if after embedding it still remains as a valid representation. It has been shown that this technique not only increases the options for embedding by increasing number of bit-planes but also gives less distortion than classical binary and Fibonacci Decomposition, while embedding message in higher bit-planes [2].

Rest of the paper is organized as follows. In Section 2, the proposed natural decomposition technique is introduced followed by experiment results in Section 4. Some conclusions are also provided towards the end.

## 2. Natural Number Decomposition

We define another new number system denoted as $(2, \mathrm{N}(.))$, where the weight function $\mathrm{N}(.)$ is defined as:

$$W(i) = \mathrm{N}(i) = i + 1, \forall i \in Z^+ \cup \{0\}$$

Since the weight function is composed of natural numbers, we name this number system as natural number system and the decomposition as natural number decomposition. In this number system, we have redundancy too. To make our transformation one-to-one, we again take the lexicographically highest of all the presentations in our number system, corresponding to same value. (e.g., value '3' has two different representations in 3-bit natural number system, namely, 100 and 011, since

$$1.3 + 0.2 + 0.1 = 3, and, 0.3 + 1.2 + 1.1 = 3$$

Since 100 is lexicographically (from left to right) higher than 011, we choose 100 to be valid representation for 3 in our natural number system and thus discard 011, no longer a valid representation in our number system.

$$3 \equiv \max_{lexicogaphic}(100,011) \equiv 100$$

In our example, the valid representations are:

$$000 \leftrightarrow 0, 001 \leftrightarrow 1, 010 \leftrightarrow 2, 100 \leftrightarrow 3,$$

$$101 \leftrightarrow 4, 110 \leftrightarrow 5, 111 \leftrightarrow 6$$

Also, to avoid loss of message, we embed secret data bit to only those pixels, where, after embedding we get a valid representation in the number system. It's worth noticing that, up-to 3-bits, the prime [2] and the natural number system are identical, after that they are different.

### 2.2. Embedding algorithm

First, we need to find a number $n \in \mathrm{N}$ such that all possible pixel values in the range $[0, 2^k - 1]$ can be represented using first *n* natural numbers (as weights) in our *n*-bit natural number system, and we get n virtual bit-planes after decomposition. To find '*n*' is quite easy, since we see, and we shall prove shortly that, in n-bit Natural Number System, all (and only) the numbers in the range $[0, n(n+1)/2]$ can be represented. So, our job reduces to finding an *n* such that $n(n+1)/2 \geq 2^k - 1$, solving the following quadratic in-equality:

$$n^2 + n - 2^{k+1} + 2 \geq 0,$$

$$\Rightarrow n \geq \frac{-1 + \sqrt{2^{k+3} + 9}}{2}, \because n \in \mathrm{Z}^+ .$$

After finding *n*, we create a map of k-bit (classical binary) to n-bit numbers (natural number decomposition), $n > k$, marking all the valid representations in our natural number system. For an 8-bit image, part of the map is illustrated in Figure-3. For k = 8, we get,

$$n \geq \frac{-1 + \sqrt{2^{8+3} + 9}}{2} = \frac{-1 + \sqrt{2057}}{2} >$$

$$\frac{-1 + 45.35}{2} = \frac{44.35}{2} = 22.675 \Rightarrow n = 23$$

Hence, we get 23 (virtual) bitplanes. Next, for each pixel of the cover image, we choose a (virtual) bit plane, say $p^{th}$ bit-plane and embed the secret data bit into that particular bit plane, by replacing the corresponding bit by the data bit, if and only if we find that after embedding the data bit, the resulting sequence is a valid representation in n-bit natural number system, i.e., exists in map – otherwise we don't hide data in that particular pixel. After embedding the secret message bit, we convert the resultant sequence in natural number system back to its value (in classical binary) and we get our stego-image. This reverse conversion is easy, since we need to calculate $\sum_{i=0}^{n-1} b_i.(i+1)$ only, where

$$b_i \in \{0,1\} \forall i \in \{0, n-1\}.$$

| N | Natural Decomp | N | Natural Decomp |
|---|---|---|---|
| 0 | 00000000000000000000000 | 32 | 10000000000000100000000 |
| 1 | 00000000000000000000001 | 33 | 10000000000001000000000 |
| 2 | 00000000000000000000010 | 34 | 10000000000010000000000 |
| 3 | 00000000000000000000100 | 35 | 10000000000100000000000 |
| 4 | 00000000000000000001000 | 36 | 10000000001000000000000 |
| 5 | 00000000000000000010000 | 37 | 10000000010000000000000 |
| 6 | 00000000000000000100000 | 38 | 10000000100000000000000 |
| 7 | 00000000000000001000000 | 39 | 10000001000000000000000 |
| 8 | 00000000000000010000000 | 40 | 10000010000000000000000 |
| 9 | 00000000000000100000000 | 41 | 10000100000000000000000 |
| 10 | 00000000000001000000000 | 42 | 10001000000000000000000 |
| 11 | 00000000000010000000000 | 43 | 10010000000000000000000 |
| 12 | 00000000000100000000000 | 44 | 10100000000000000000000 |
| 13 | 00000000001000000000000 | 45 | 11000000000000000000000 |
| 14 | 00000000010000000000000 | 46 | 11000000000000000000001 |
| 15 | 00000000100000000000000 | 47 | 11000000000000000000010 |
| 16 | 00000001000000000000000 | 48 | 11000000000000000000100 |
| 17 | 00000010000000000000000 | 49 | 11000000000000000001000 |
| 18 | 00000100000000000000000 | 50 | 11000000000000000010000 |
| 19 | 00001000000000000000000 | 51 | 11000000000000000100000 |
| 20 | 00010000000000000000000 | 52 | 11000000000000001000000 |
| 21 | 00100000000000000000000 | 53 | 11000000000000010000000 |
| 22 | 01000000000000000000000 | 54 | 11000000000000100000000 |
| 23 | 10000000000000000000000 | 55 | 11000000000001000000000 |
| 24 | 10000000000000000000001 | 56 | 11000000000010000000000 |
| 25 | 10000000000000000000010 | 57 | 11000000000100000000000 |
| 26 | 10000000000000000000100 | 58 | 11000000001000000000000 |
| 27 | 10000000000000000001000 | 59 | 11000000010000000000000 |
| 28 | 10000000000000000010000 | 60 | 11000000100000000000000 |
| 29 | 10000000000000000100000 | 61 | 11000001000000000000000 |
| 30 | 10000000000000001000000 | 62 | 11000010000000000000000 |
| 31 | 10000000000000010000000 | 63 | 11000100000000000000000 |

**Figure-1.** Natural number decomposition for 8-bit image yielding 23 virtual bit-planes, whereas prime decomposition yields much less number of virtual bit planes [2], as depicted in Figure 2.

| N | Prime Decomp. | N | Prime Decomp. | N | Prime Decomp. |
|---|---|---|---|---|---|
| 0 | 000000000000000 | 32 | 000100000000001 | 64 | 100000100000010 |
| 1 | 000000000000001 | 33 | 000100000000010 | 65 | 100000100000100 |
| 2 | 000000000000010 | 34 | 000100000000100 | 66 | 100001000000000 |
| 3 | 000000000000100 | 35 | 000100000000101 | 67 | 100001000000001 |
| 4 | 000000000000101 | 36 | 000100000001000 | 68 | 100001000000010 |
| 5 | 000000000001000 | 37 | 001000000000000 | 69 | 100001000000100 |
| 6 | 000000000001001 | 38 | 001000000000001 | 70 | 100001000000101 |
| 7 | 000000000010000 | 39 | 001000000000010 | 71 | 100001000001000 |
| 8 | 000000000010001 | 40 | 001000000000100 | 72 | 100010000000000 |
| 9 | 000000000010010 | 41 | 010000000000000 | 73 | 100010000000001 |
| 10 | 000000000010100 | 42 | 010000000000001 | 74 | 100100000000000 |
| 11 | 000000000100000 | 43 | 100000000000000 | 75 | 100100000000001 |
| 12 | 000000000100001 | 44 | 100000000000001 | 76 | 100100000000010 |
| 13 | 000000001000000 | 45 | 100000000000010 | 77 | 100100000000100 |
| 14 | 000000001000001 | 46 | 100000000000100 | 78 | 100100000000101 |
| 15 | 000000001000010 | 47 | 100000000000101 | 79 | 100100000001000 |
| 16 | 000000001000100 | 48 | 100000000001000 | 80 | 101000000000000 |
| 17 | 000000010000000 | 49 | 100000000001001 | 81 | 101000000000001 |
| 18 | 000000010000001 | 50 | 100000000010000 | 82 | 101000000000010 |
| 19 | 000000100000000 | 51 | 100000000010001 | 83 | 101000000000100 |
| 20 | 000000100000001 | 52 | 100000000010010 | 84 | 110000000000000 |
| 21 | 000000100000010 | 53 | 100000000010100 | 85 | 110000000000001 |
| 22 | 000000100000100 | 54 | 100000000100000 | 86 | 110000000000010 |
| 23 | 000001000000000 | 55 | 100000000100001 | 87 | 110000000000100 |
| 24 | 000001000000001 | 56 | 100000001000000 | 88 | 110000000000101 |
| 25 | 000001000000010 | 57 | 100000001000001 | 89 | 110000000001000 |
| 26 | 000001000000100 | 58 | 100000001000010 | 90 | 110000000001001 |
| 27 | 000001000000101 | 59 | 100000001000100 | 91 | 110000000010000 |
| 28 | 000001000001000 | 60 | 100000010000000 | 92 | 110000000010001 |
| 29 | 000010000000000 | 61 | 100000010000001 | 93 | 110000000010010 |
| 30 | 000010000000001 | 62 | 100000100000000 | 94 | 110000000010100 |
| 31 | 000100000000000 | 63 | 100000100000001 | 95 | 110000000100000 |

**Figure-2**. Prime decomposition for 8-bit image yielding 15 virtual bit-planes

### 2.3. Extraction algorithm

The extraction algorithm is exactly the reverse. From the stego-image, each pixel with embedded data bit is converted to its corresponding natural decomposition and from the $p^{th}$ bit-plane the secret message bit is extracted. Further, all the bits are combined to get the secret message.

### 2.4. Performance analysis–comparison of Prime & Natural Decomposition

**Lemma-1**. In k-bit Natural Number System, all numbers in the range $[0, k(k+1)/2]$ can be represented and only these numbers can be represented. (Proof: simply use induction on k)

**Lemma-2.** The Natural Decomposition generates more (virtual) bit-planes

$\lim_{n\to\infty} \frac{p_n}{n\ln(n)} = 1$ (by Prime Number Theorem), if $p_n$ be the $n^{th}$ prime $\Rightarrow p_n = \theta(n.\ln(n))$ [11]

$\because n+1 = o(n.\ln(n))$, the weight corresponding to the n[th] bit in our number system using natural number decomposition eventually becomes much higher than the weight corresponding to the n[th] bit in the number system using prime decomposition. In n-bit Prime Number System, numbers in the range $[0, \sum_{i=0}^{n-1} p_i]$ can be represented, while in n-bit Natural Number System, numbers in the range $[0, \sum_{i=0}^{n-1}(i+1)] = [0, \sum_{i=1}^{n} i] = [0, n(n+1)/2]$ can be represented. Now, it's easy to prove that $\exists n_0 \in \mathrm{N} : \forall n \geq n_0$, we have,

$\sum_{i=0}^{n-1} p_i > \frac{n(n+1)}{2}$.

So, using same number of bits it is eventually possible to represent more numbers in case of natural number decomposition than in case of prime decomposition. This implies that number of (virtual) bit-planes generated for natural number decomposition will be eventually more than that of prime decomposition.

**Lemma-3.** The Natural Decomposition gives less distortion in higher bit-planes. (Here, we assume the secret message length is same as image size. For message with different length, the same can similarly be derived in a straight-forward manner) As before, we use Worst-case-Mean-Square-Error (WMSE) and the corresponding PSNR (per pixel) as our test-statistics. In case of Prime Decomposition:

$\left(WMSE_{l^{th} bit-plane}\right)_{\mathrm{Prime-Decomposition}} = w \times h \times p_l^2 = \theta(l^2.\log^2(l))$.

In case of our Natural Decomposition, WMSE for embedding secret message bit only in $l^{th}$ (virtual) bit-plane of each pixel (after expressing a pixel in our natural number system, using natural number decomposition technique) $= (l+1)^2$.

$\therefore \left(WMSE_{l^{th} bit-plane}\right)_{Natural-Number-Decomposition}$

$= w \times h \times (l+1)^2 = \theta(l^2)$.

$\because (l+1)^2 = o(l^2.\log^2(l))$, eventually we have,

$\left(WMSE_{l^{th} bit-plane}\right)_{Natural-Number-Decomposition}$

$< \left(WMSE_{l^{th} bit-plane}\right)_{\mathrm{Prime-Decomposition}}$

The above result implies that the distortion in case of natural number decomposition is much less than that in case of prime decomposition. Figure-3 illustrates our claim and it compares the nature of the weight function in case of prime decomposition against that of natural number decomposition.

$$\left(WMSE_{l^{th} bit-plane}\right)_{Classical-Binary-Decomposition} = \theta(4^l).$$

$$\left(WMSE_{l^{th} bit-plane}\right)_{\mathrm{Prime-Decomposition}} = \theta(l^2.\log^2(l)).$$

$$\left(WMSE_{l^{th} bit-plane}\right)_{Natural-Number-Decomposition} = \theta(l^2).$$

- $(PSNR_{worst})_{Classical-Binary-Decomposition}$

$$= 10.\log_{10}\left(\frac{(2^k-1)^2}{(2^l)^2}\right).$$

- $(PSNR_{worst})_{\mathrm{Prime-Decomposition}}$

$$= 10.\log_{10}\left(\frac{(2^k-1)^2}{(c.l^2.\log^2(l))^2}\right), c \in \Re^+.$$

- $(PSNR_{worst})_{Natural-Number-Decomposition}$

$$= 10.\log_{10}\left(\frac{(2^k-1)^2}{(l+1)^2}\right).$$

**Lemma-4.** Natural Decomposition is Optimal

This particular decomposition technique is optimal in the sense that it generates maximum number of (virtual) bit-planes and also least distortion while embedding in higher bit-planes, when the weight function is monotonically strictly increasing. Since among all monotonic strictly increasing sequences of positive integers, natural number sequence is the tightest, all others are subsequences of the natural number sequence. Since we have the weight function $W : \mathrm{Z}^+ \cup \{0\} \to \mathrm{Z}^+$, that assigns a bit-plane (index) an integral weight, if we assume that weight corresponding to a bit-plane is unique and the weight is monotonically increasing, one of the simplest but yet optimal way to construct such an weight function is to assign consecutive natural number values to the weights corresponding to each bit-plane, i.e., (We defined the value of $W(i) = i + 1$ instead of $W(i) = i$, since we want all-zero representation for the value 0, in this number system).

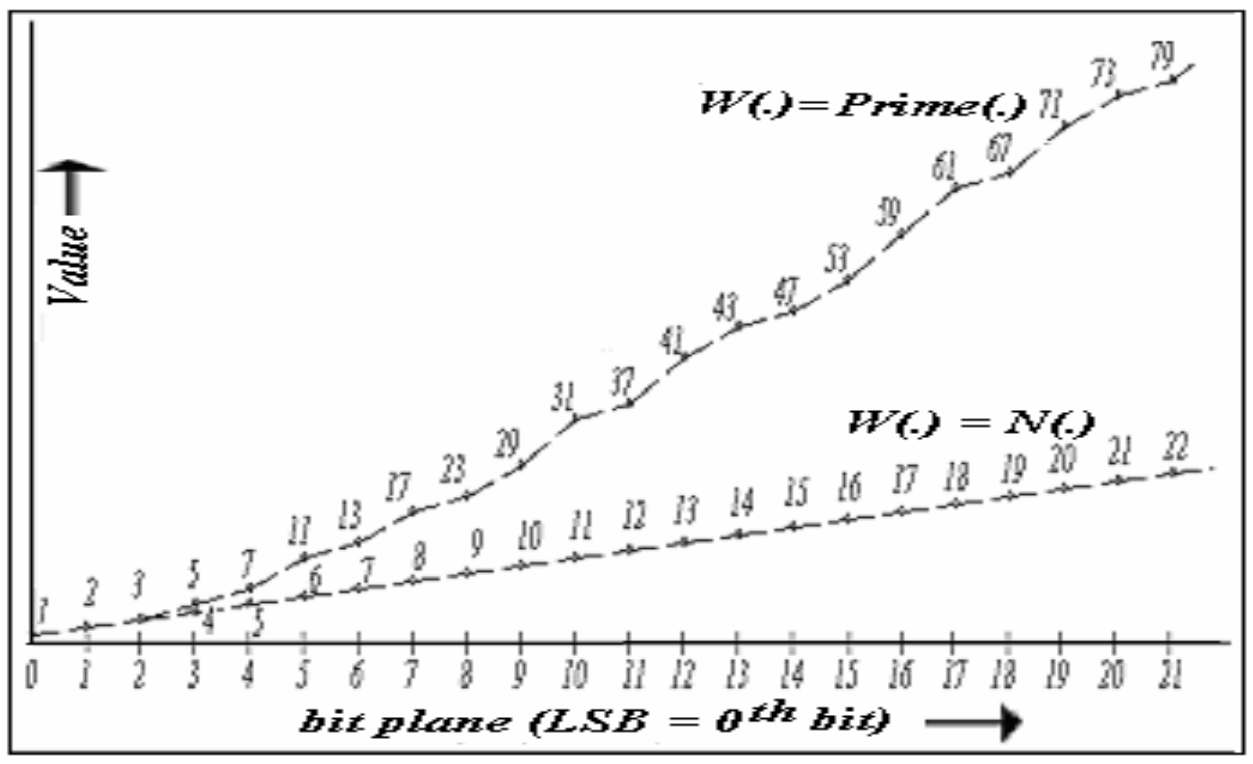

**Figure-3.** Weight functions for different decomposition techniques

Now, this particular decomposition in virtual bit-planes and embedding technique gives us optimal result. We get optimal performance of any data-hiding technique by minimizing our test-statistic WMSE. For embedding data in $l^{th}$ virtual bit-plane, we have,

$(WMSE)_{l^{th} bit-plane} = (W(l))^2$, so minimizing WMSE implies minimizing the weight function $W(.)$, but having our weight function allowed to assume integral values only, and also assuming the values assigned by W are unique (W is injective, we discard the un-interesting case when weight-values corresponding to more than one bit-planes are equal), we can without loss of generality assume W to be monotonically increasing. But, according to the above condition imposed on W, we see that such strictly increasing W assigning minimum integral weight-values to different bit planes must be linear in bit-plane index.

Put it in another way, for n-bit number system, we need n different weights that are to be assigned to weight-values corresponding to n bit-planes. But, the assigning must also guarantee that these weight values are minimum possible. Such n different positive integral values must be smallest n consecutive natural numbers, i.e., 1, 2, 3, …, n. But, our weight function $W(i) = i + 1, \forall i \in Z^+ \cup \{0\}$ merely gives these values as weights only, hence this technique is optimal.

Using classical binary decomposition, we get k bit planes only corresponding to a k-bit pixel value, but for natural number decomposition, we get, n bit pixels, where n satisfies,

$$n^2 + n - 2^{k+1} + 2 \geq 0,$$

$$\Rightarrow n \geq \frac{-1 + \sqrt{2^{k+3} + 9}}{2}, \because n \in Z^+ \Rightarrow n = \theta(2^{k/2})$$

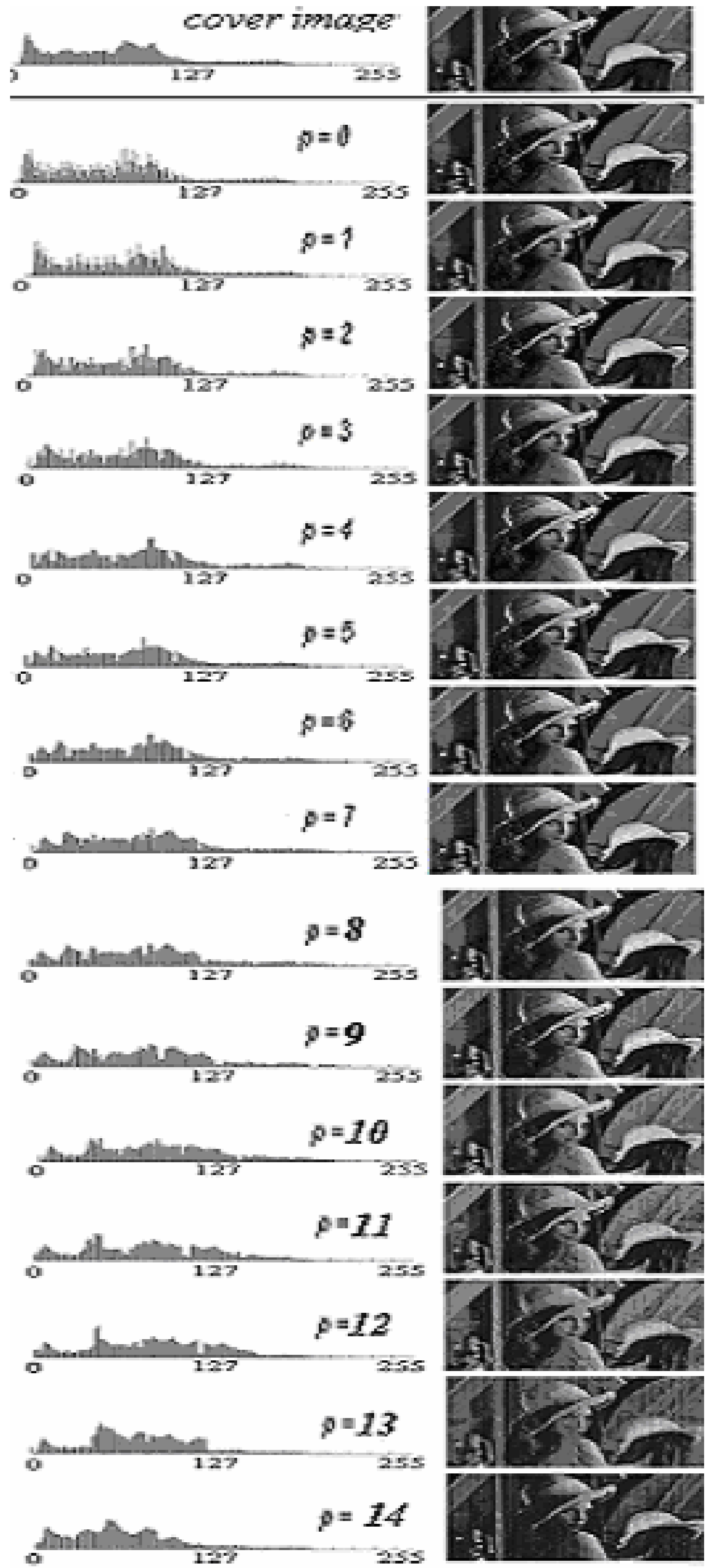

**Figure-4**. Frequency distribution of pixel gray levels in different bit-planes before and after data-hiding in case of Prime decomposition

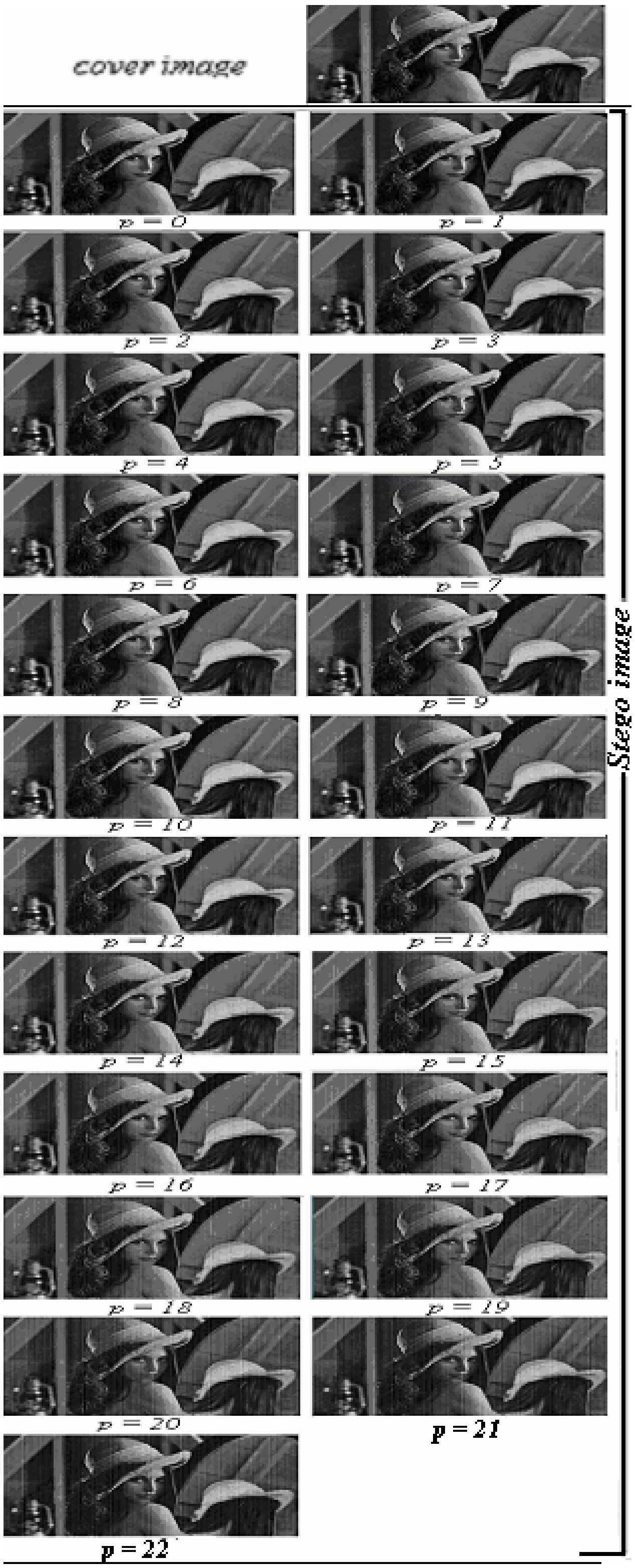

**Figure-5**. Embedding data in different bit-planes using Natural Number decomposition

## 3. Experimental Results for Natural Number decomposition technique

As input the 8-bit gray-level image of Lena is used. We set the secret message length = cover image size, (message string '*sandipan*' repeated multiple times to fill the cover image size). The secret message bits are embedded into chosen bit-plane '*p*' using different decomposition techniques, namely, the classical binary (LSB), Fibonacci, Prime and Natural Number decomposition separately and compared.

Figures 4 and 5 illustrate that, we get 15 and 23 bit-planes for Prime and Natural Number decomposition techniques respectively and the change of frequency distribution corresponding to gray-level values is least in case of Natural when compared to the other techniques, eventually resulting in a still less relative entropy between the cover-image and stego-image, implying least visible distortions, as we move towards higher bit-planes for embedding data bits. This technique can also be enhanced by embedding into more than one (virtual) bit-plane, following the variable-depth data-hiding technique [6].

## 4. Conclusions

This paper presented very simple method of data hiding technique using natural numbers. It is shown (both theoretically and experimentally) that the data-hiding technique natural number decomposition outperforms the one using prime, Fibonacci and classical LSB decomposition, in terms of embedding secret data bits at higher bit-planes with less detectable distortion. We have shown all our experimental results using the famous Lena image, but since in all our theoretical derivation above we have shown our test-statistic value (WMSE, PSNR) independent of the probability mass function of the gray levels of the input image, the (worst-case) result will be similar if we use any gray-level image as input, instead of the Lena image.